\documentclass[preprint,nofootinbib]{revtex4}%
\usepackage{amssymb}
\usepackage{amsfonts}
\usepackage{amsmath}
\usepackage{graphicx}
\usepackage[usenames]{color}%
\providecommand{\U}[1]{\protect\rule{.1in}{.1in}}
\definecolor{blue}{rgb}{0,0,1}

\definecolor{red}{rgb}{1,0,0}

\begin{document}
\title{Black Strings in Gauss-Bonnet Theory are Unstable}
\author{Alex Giacomini$^{1}$, Julio Oliva$^{1}$, Aldo Vera$^{2}$}
\affiliation{$^{1}$Instituto de Ciencias F\'{\i}sicas y Matem\'{a}ticas, Universidad
Austral de Chile, Casilla 567, Valdivia, Chile}
\affiliation{$^{2}$Departamento de F\'{\i}sica, Universidad de Concepci\'{o}n, Casilla
160-C, Concepci\'{o}n, Chile}

\begin{abstract}
We report the existence of unstable, s-wave modes, for black strings in
Gauss-Bonnet theory (which is quadratic in the curvature) in seven dimensions.
This theory admits analytic uniform black strings that in the transverse
section are black holes of the same Gauss-Bonnet theory in six dimensions. All
the components of the perturbation can be written in terms of a single one and
its derivatives. For this latter component we find a master equation which
admits bounded solutions provided the characteristic time of the exponential
growth of the perturbation is related with the wave number along the extra
direction, as it occurs in General-Relativity. It is known that these
configurations suffer from a thermal instability, and therefore the results
presented here provide evidence for the Gubser-Mitra conjecture in the context
of Gauss-Bonnet theory. Due to the non-triviality of the curvature of the
background, all the components of the metric perturbation appear in the
linearized equations. As it occurs for spherical black holes, these black
strings should be obtained as the short distance limit $r<<\alpha^{1/2}$ of
the black string solution of Einstein-Gauss-Bonnet theory, which is not know
analytically, where $\alpha$ is the Gauss-Bonnet
coupling.\footnote{alexgiacomini@uach.cl, julio.oliva@uach.cl,
aldovera@udec.cl}

\end{abstract}
\maketitle

\section{Introduction}

Gravity in higher dimensions has been an important scenario to test how
generic are the notions we have gained from four-dimensional gravity. Also
motivated by String Theory and supergravity, many results have been learned in
the last decades concerning gravity in dimensions higher than four, as for
example the existence of the asymptotically flat black rings and all its
extensions (for a review see \cite{Emparan:2008eg} and \cite{horowitzbook}).
These objects were conjectured to be unstable for large angular momentum, as
they inherit the Gregory-Laflamme instability \cite{Emparan:2001wn} of
non-extremal black strings and black p-branes \cite{Gregory:1993vy}%
-\cite{Gregory:1994bj}. Indeed the black ring instability has been confirmed
in \cite{Hovdebo:2006jy}, \cite{Figueras:2011he}, \cite{Santos:2015iua}. The
Gregory-Laflamme instability can be guessed from thermodynamical arguments
since, as a function of the mass, the entropies of the black hole and the black
string cross at a given critical mass $M_{c}$. This can be seen from the fact
that the entropy of the black hole grows as $S_{BH}\sim M^{\frac{D-2}{D-3}}$,
while the entropy of the black string grows as $S_{BS}\sim M^{\frac{D-3}{D-4}%
}$. For masses below $M_{c}$, the black hole is thermally favoured and above
$M_{c}$ is the black string solution the one with greater entropy and
therefore the most favoured. This relation between thermal and perturbative
instabilities led Gubser and Mitra to conjecture that both kinds of
instabilities always appear together for black hole configurations with
extended directions \cite{Gubser:2000mm}, and is was recently proved in
\cite{Hollands:2012sf} for General Relativity in vacuum. To understand the
complete evolution of the unstable mode a non-linear analysis is required.
Recent outstanding numerical results in five dimensions seem to indicate that the black string
evolves toward a non-homogenous configuration with section for which the size
of the string eventually shrinks to zero generating a null singularity and
providing a counterexample of the cosmic censorship conjecture
\cite{Lehner:2010pn} (for a historical review on this problem see Chapter 2 of \cite{horowitzbook}).

An interesting problem is whether higher curvature corrections may modify this
scenario. In the particular case of higher curvature Lovelock theories
\cite{Lovelock:1971yv} it is difficult to construct analytic, homogenous,
black strings due to the fact that the new dimensionful coupling constants
introduce a length scale that induces the existence of a cosmological
constant. Numerical and approximate results in this context have been reported
in \cite{Barcelo:2002wz}-\cite{Kobayashi:2004hq}. Specially, in \cite{Brihaye:2010me}, static uniform and non-uniform black strings where constructed. The latter were constructed along the lines of \cite{Gubser:2001ac}, i.e., perturbativelly in a non-homogeneity parameter and therefore can be considered as a static perturbation of the uniform black string. Comparing the entropies of these configurations the authors provided evidence for the Gubser and Mitra conjecture in the context of Einstein-Gauss-Bonnet theory.

The situation in theories that have a single Lovelock term is much more like the one in General
Relativity, since as shown in \cite{Giribet:2006ec} homogeneous black strings
and black p-branes can be constructed analytically. These solutions are also
important since they should be obtained as the short distance configuration
($r<<\alpha^{1/2}$) of the black string solution of Einstein-Gauss-Bonnet
theory, which is not know analytically. Here $\alpha$ is the Gauss-Bonnet
coupling. This is what occurs for example with the ``healthy branch'' of the
spherically symmetric black hole in Einstein-Gauss-Bonnet gravity, which is
defined by the following action:%
\begin{equation}
I_{EGB}\left[  g\right]  =\frac{1}{16\pi G}\int d^{D}x\sqrt{-g}\left[
R+\alpha\left(  R^{2}-4R_{\mu\nu}R^{\mu\nu}+R_{\alpha\beta\gamma\delta
}R^{\alpha\beta\gamma\delta}\right)  \right]  \ .
\end{equation}

This theory admits a the following black hole solution \cite{Boulware:1985wk}:%
\begin{equation}
ds^{2}=-f\left(  r\right)  dt^{2}+\frac{dr^{2}}{f\left(  r\right)  }%
+r^{2}d\Omega_{D-2}^{2}\ ,
\end{equation}
where%
\begin{equation}
f\left(  r\right)  =1+\frac{r^{2}}{\left(  D-3\right)  \left(  D-4\right)
\alpha}\left[  1-\sqrt{1+\frac{64\pi G\left(  D-3\right)  \left(  D-4\right)
\alpha}{\left(  D-2\right)  V\left(  S^{D-2}\right)  }\frac{M}{r^{D-1}}%
}\right]  \ .
\end{equation}
where the integration constant $M$, is the mass. Here $\alpha$ has dimensions
of length square and we can analyze the behavior of this metric function for
$r>>\sqrt{\alpha}$ and $r<<\sqrt{\alpha}$ which respectively read:%
\begin{align}
&  f\left(  r\right)  \underset{r>>\sqrt{\alpha}}{\approx}1-\frac{32\pi
G}{\left(  D-2\right)  V\left(  S^{D-2}\right)  }\frac{M}{r^{D-3}%
}+...\label{finf}\\
&  f\left(  r\right)  \underset{r<<\sqrt{\alpha}}{\approx}1-\left(
\frac{64\pi GM}{\left(  D-3\right)  \left(  D-4\right)  \left(  D-2\right)
\alpha V\left(  S^{D-2}\right)  }\right)  ^{\frac{1}{2}}\frac{1}{r^{\frac
{D-5}{2}}}+...\ . \label{fzero}%
\end{align}
In the former case the solution reduces to the Schwarzschild-Tangherlini black
hole, while in the latter it reduces to the solution found in
\cite{Crisostomo:2000bb}. Therefore, we have that for large distances, the
effects of the quadratic curvature term are sub-leading, whereas for
short distance (as compared with $\sqrt{\alpha}$) the quadratic terms dominate
and one recovers a solutions of Gauss-Bonnet theory.

As shown in \cite{Giribet:2006ec}, the asymptotically flat black holes
constructed in \cite{Crisostomo:2000bb} can be oxidated to construct
homogenous black string and black p-brane solutions. These spacetimes are
solutions of the theory that contains only the $k-th$ order term in the
Lovelock theory, being the case $k=1$ the one of General Relativity. For
simplicity let's consider only the quadratic Gauss-Bonnet term in seven
dimension:%
\begin{equation}
I_{EGB}\left[  g\right]  =\frac{\alpha}{16\pi G}\int d^{7}x\sqrt{-g}\left[
R^{2}-4R_{\mu\nu}R^{\mu\nu}+R_{\alpha\beta\gamma\delta}R^{\alpha\beta
\gamma\delta}\right]  \ .
\end{equation}
This theory has the following two solution%
\begin{equation}
ds^{2}=-\left(  1-\frac{\mu}{r}\right)  dt^{2}+\frac{dr^{2}}{1-\frac{\mu}{r}%
}+r^{2}d\Omega_{5}^{2}\ , \label{bh2}%
\end{equation}
and
\begin{equation}
ds^{2}=-\left(  1-\frac{m}{r^{1/2}}\right)  dt^{2}+\frac{dr^{2}}{1-\frac
{m}{r^{1/2}}}+r^{2}d\Omega_{4}^{2}+dz^{2}\ , \label{bs2}%
\end{equation}
which correspond to an spherically symmetric black hole and a black string,
respectively. The constants $m$ and $\mu$ determine the masses of the
configurations while $d\Omega_{n}$ stands for the line element of the
$n-$sphere, $S^{n}$. From the experience gained from the spherically symmetric
black hole, one can expect that these black strings should be obtained as the
short distance limit of the, not known analytically, black string of
Einstein-Gauss-Bonnet theory in seven dimensions.

The black strings and black p-branes constructed in this way where proved to
be thermally unstable \cite{Giribet:2006ec} exactly in the same manner than
the black strings in General Relativity, since the entropies, as a function of
the mass for (\ref{bh2}) and (\ref{bs2}) read $S_{BH}^{GB}\sim M^{\frac{3}{2}%
}$ and $S_{BS}^{GB}\sim M^{2}$, respectively and they do cross at a critical
mass $M_{c}^{GB}$. The heat capacities of the black hole (\ref{bh2}) and the black hole in the transverse section of (\ref{bs2}) are negative (see \cite{Crisostomo:2000bb}), therefore, both black objects (\ref{bh2}) and (\ref{bs2}) are locally thermally unstable\footnote{The temperature of the black string (\ref{bs2}) is the same as that of the black hole on its transverse section and the mass of such black string corresponds to the mass of the black hole times the extension of the extended direction. Therefore the sign of the heat capacity remains the same after the oxidation.}. 

A natural question is whether such thermal instability has a perturbative
counterpart. In this paper we show this is indeed the case. In the next
section we show that the black strings of Gauss-Bonnet theory (\ref{bs2}) are
unstable under the s-wave mode and that such instability disappears for
compactified black strings that are short enough.

\section{The perturbative instability}

Here we will be concerned with gravitational perturbations in the context of
Gauss-Bonnet theory. The field equations are therefore given by%
\begin{equation}
E_{\mu\nu}:=2RR_{\mu\nu}-4R_{\mu\rho\nu\sigma}R^{\rho\sigma}+2R_{\mu\rho
\sigma\tau}R_{\nu}^{\ \rho\sigma\tau}-4R_{\mu\rho}R_{\nu}^{\ \rho}-\frac{1}%
{2}g_{\mu\nu}\left(  R^{2}-4R_{\alpha\beta}R^{\alpha\beta}+R_{\alpha
\beta\gamma\delta}R^{\alpha\beta\gamma\delta}\right)  =0\ . \label{feq}%
\end{equation}
For simplicity we will focus in the seven dimensional case. As mentioned above
, this theory admits the homogenous black string solution (\ref{bs2}). The
radius of the horizon reads $r_{+}=m^{2}$. In order to work with a finite
range of parameters, let's consider the change in the radial coordinate given
by%
\begin{equation}
r=\left(  \frac{m}{1-x}\right)  ^{2}\ ,
\end{equation}
that maps the region outside the event horizon $r\in\lbrack m^{2}%
,+\infty\lbrack$ to $x\in\lbrack0,1[\ $. In this new coordinates the metric
(\ref{bs2}) reads%
\begin{equation}
ds_{BS_7}^{2}=-xdt^{2}+\frac{4m^{4}dx^{2}}{x\left(  1-x\right)  ^{6}}+\left(
\frac{m}{1-x}\right)  ^{2}d\Omega_{4}^{2}+dz^{2}\ . \label{bsx}%
\end{equation}
The s-wave perturbation on the background, black string metric (\ref{bsx})
reads%
\[
h_{\mu\nu}\left(  t,x,z\right)  =e^{\Omega t}e^{ikz}\left(
\begin{tabular}
[c]{cc|c|c}%
$H_{tt}\left(  x\right)  $ & $H_{tx}\left(  x\right)  $ & $0$ & $0$\\
$H_{tx}\left(  x\right)  $ & $H_{xx}\left(  x\right)  $ & $0$ & $0$\\\hline
$0$ & $0$ & $H\left(  x\right)  \sigma_{S^{4}}$ & $0$\\\hline
$0$ & $0$ & $0$ & $0$%
\end{tabular}
\ \right)  \ ,
\]
where $\sigma_{S^{4}}$ is the metric of the four sphere and $k$ is the wave
number along the $z$ direction. An unstable mode is defined as a bounded
solution of the linearized Gauss-Bonnet equations (\ref{feq}) with positive
$\Omega$. It easy to show that the linearized field equations imply that the
components of the perturbation can be written in terms of $H_{tx}\left(
x\right)  $ in the following manner%
\begin{align}
H_{tt}\left(  x\right)   &  =\frac{\left(  1-x\right)  ^{6}x^{2}}{4m^{4}%
\Omega^{4}}H_{tx}^{\prime}+\frac{x\left(  1-x\right)  ^{6}}{4m^{4}\Omega
}H_{tx}\label{ttdetx}\\
H_{xx}\left(  x\right)   &  =-\frac{x}{\Omega}H_{tx}^{\prime\prime}%
-\frac{2\left(  1-4x\right)  }{\left(  1-x\right)  \Omega}H_{tx}^{\prime
}+\left(  \frac{\left(  3k^{2}x+4\Omega^{2}\right)  m^{4}}{x\left(
1-x\right)  ^{6}\Omega}+\frac{6}{\Omega\left(  1-p\right)  }\right)
H_{tx}\label{xxdetx}\\
H\left(  x\right)   &  =\frac{x^{2}\left(  1-x\right)  ^{2}}{6\Omega}%
H_{tx}^{\prime\prime}+\frac{\left(  1-3x\right)  \left(  1-x\right)
x}{2\Omega}H_{tx}^{\prime}+\left(  \frac{\left(  1-x\right)  \left(
1-7x\right)  }{6\Omega}-\frac{m^{4}\left(  3k^{2}x+4\Omega^{2}\right)
}{6\left(  1-x\right)  ^{4}\Omega}\right)  H_{tx}, \label{hdetx}%
\end{align}
where the prime ($^{\prime}$) denotes differentiation with respect to $x$. The
component $H_{tx}\left(  x\right)  $ fulfils the following linear, second
order, master equation \begin{equation}
A\left(  x\right)  H_{tx}^{\prime\prime}+B\left(  x\right)  H_{tx}^{\prime
}+C\left(  x\right)  H_{tx}=0\ , \label{mastereq}%
\end{equation}
with%
\begin{align}
A\left(  x\right)   &  =(1-x)^{6}x^{2}\left(  (1-x)^{6}-(12k^{2}x+16\Omega
^{2})m^{4}\right)  \ ,\\
B\left(  x\right)   &  =3x(1-x)^{5}\left(  (32k^{2}x^{2}+48x\Omega^{2}%
-8k^{2}x-16\Omega^{2})m^{4}+(1-x)^{7}\right)  \ ,\\
C\left(  x\right)   &  =4\left(  4\Omega^{2}+3k^{2}x\right)  ^{2}%
m^{8}+(1-x)^{5}(45k^{2}x^{2}+164x\Omega^{2}+3k^{2}x-20\Omega^{2}%
)m^{4}+(1-x)^{12}\ .
\end{align}
Then, one can see that all the linearized field equations are solved provided
(\ref{ttdetx})-(\ref{hdetx}) and (\ref{mastereq}), hold. Note that the master
equation is invariant under%
\begin{equation}
m\rightarrow\gamma m\text{, }\Omega\rightarrow\gamma^{-2}\Omega\text{,
}k\rightarrow\gamma^{-2}k\ , \label{scalingsymmetry}%
\end{equation}
for an arbitrary constant $\gamma$. Therefore it is enough to study the existence of unstable modes for a fixed
value of the horizon radius $r_{+}=m^{2}$, since the other can be obtained by
applying the scaling symmetry (\ref{scalingsymmetry}).

We are then left with finding a well-behaved solution of the master equation
(\ref{mastereq}). This equation implies that the solution $H_{tx}\left(
x\right)  $ admits the following asymptotic behaviors at the horizon
$(x\rightarrow0)$ and at infinity $\left(  x\rightarrow+1\right)  $,
respectively:%
\begin{align}
&  H_{tx}\underset{x\rightarrow0}{\rightarrow}C_{\pm}x^{-1\pm2m^{2}\Omega
}\left(  1+\mathcal{O}\left(  x\right)  \right)  \ ,\label{asymphor}\\
&  H_{tx}\underset{x\rightarrow1}{\rightarrow}E_{\pm}\left(  1-x\right)
^{\alpha_{\pm}}e^{\mp\frac{m^{2}\sqrt{3k^{2}+4\Omega^{2}}}{2\left(
1-x\right)  ^{2}}\mp\frac{(8\Omega^{2}+3k^{2})m^{2}\sqrt{3k^{2}+4\Omega^{2}}%
}{2(3k^{2}+4\Omega^{2})\left(  1-x\right)  }}\left(  1+\mathcal{O}\left(
1-x\right)  \right)  \ , \label{asympinf}%
\end{align}
with%
\begin{equation}
\alpha_{\pm}=\frac{1}{8}\frac{-12(3k^{2}+4\Omega^{2})^{2}\pm m^{2}\left(
144k^{2}\Omega^{2}+128\Omega^{4}+27k^{4}\right)  \sqrt{3k^{2}+4\Omega^{2}}%
}{(3k^{2}+4\Omega^{2})^{2}}\ .
\end{equation}
Since we are looking for unstable modes, we need to find a numerical solution
that interpolates between the plus sign in (\ref{asymphor}) and the minus sign
in (\ref{asympinf}). It's natural to think that in order to have a well posed
behavior at the horizon we need to impose $\Omega>\Omega_{c,GB}:=\frac
{1}{2m^{2}}$, as it was originally considered in \cite{Gregory:1987nb} where
it was proved that in the five dimensional black string in General Relativity
there is no non-singular, single, unstable mode in this family (in G.R. in
five dimensions $\Omega_{c,G.R.}=\frac{1}{r_{+}}$). Nevertheless in General
Relativity, in the range $0<\Omega<\Omega_{c,GR}$ one can construct a
perturbation that is a composition of single divergent modes at the horizon in
such a manner that the divergences cancel, as it occurs for the instabilities
in some colored black holes \cite{Bizon:1991nt} and originally observed by
Vishveshwara in \cite{Vishveshwara:1970cc}. This can be seen also considering
the fact that a $t=const$ surface intersects the bifurcation surface rather
than the future horizon. It it therefore necessary consider Kruskal-like
coordinates, where the $T=const$ surfaces do indeed intersect the future horizon. Then, by going to Kruskal coordinates, it is easy to see that the
unstable modes we find below are regular at the future horizon provided we
choose the branch with the plus sign in (\ref{asymphor}), even if $\Omega<(2m^2)^{-1}$.

In order to find whether the master equation (\ref{mastereq}) admits a bounded
solution for some positive values of $\Omega$, we will follow the approach
developed in \cite{Horowitz:1999jd} for quasinormal modes. Briefly, the method
consists in proposing a power series solution around the horizon, then
selecting the well behaved branch and finally truncate the power series to
some order $N$. Then use the fact that, due to the pole structure of the
equation such a power series has a convergence radius that at least includes
$x=1$ and therefore we can request for the truncated series to vanish at
infinity ($x=1$). Such an equation provides for the spectrum of unstable
modes. For details we refer to the original work \cite{Horowitz:1999jd}.

The results of the previous analysis are depicted in Figure 1:

\begin{figure}[h]
\includegraphics[scale=0.5]{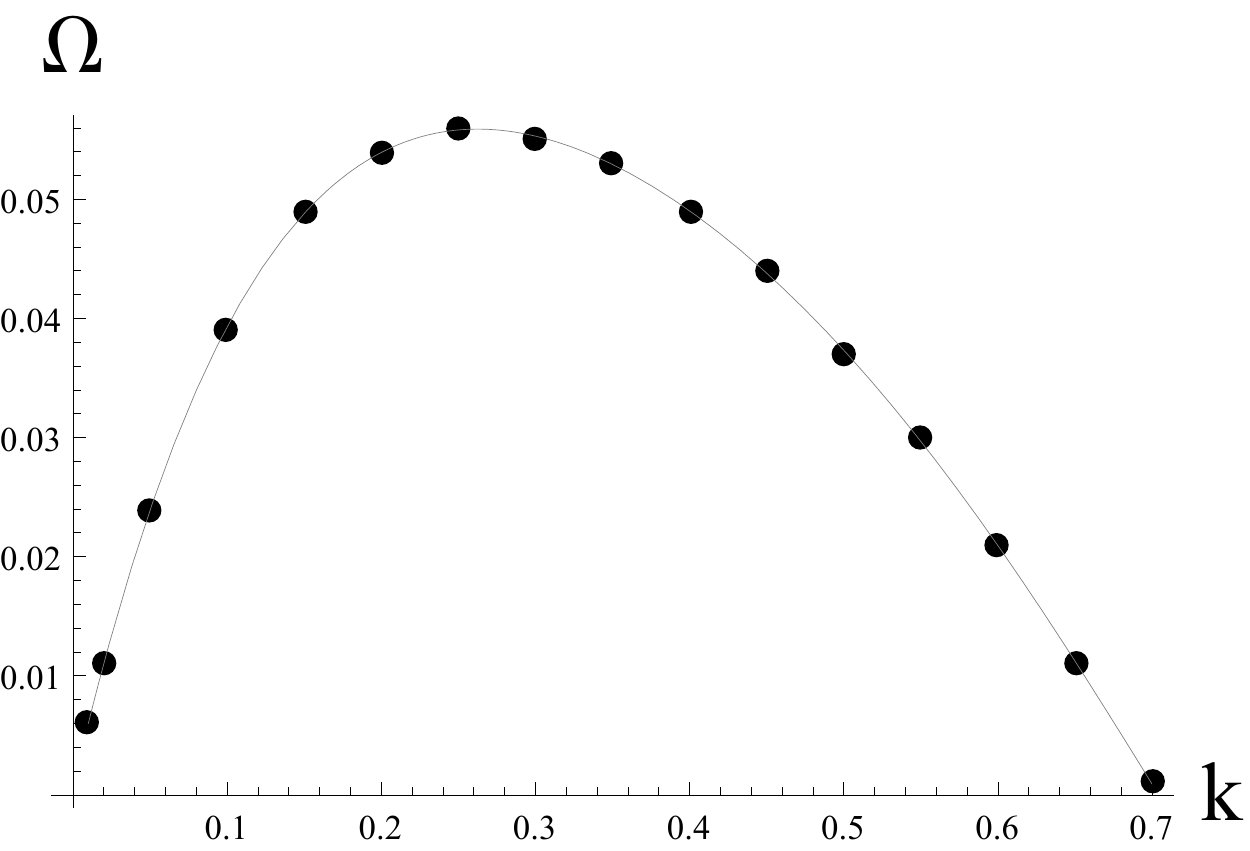} \caption{\textit{$\Omega$ vs $k$ for the
homogeneous black string in Gauss-Bonnet in $D=7$. The parameter $m$ in the
solution has been fixed to 1, and any value for the mass can be obtained by
applying the scaling transformation in (\ref{scalingsymmetry}). The numerical
precision is such that four digits in $\Omega$s are stable (the continuous
curve has been included to facilitate the visualization).}}%
\label{omegavsk}%
\end{figure}

From these results we see that there is minimum wavelength $\lambda_{min}$
above which instability occurs. This also implies the existence of a critical
length for the string, above which the instability takes place.

As it occurs for the black string in General Relativity \cite{Gregory:1994bj},
as far as $k\neq0$, one can show that the perturbation cannot be gauged away.
Another straighforward method to check that these perturbations are physical
and cannot be gauged away is to consider the following scalar invariant%
\begin{equation}
A=881R_{abcd}R_{\ \ ef}^{cd}R^{efab}+2428R_{\ \ cd}^{ab}R_{\ \ bf}%
^{ce}R_{\ \ \ ae}^{df}\ ,
\end{equation}
which vanishes identically on the unperturbed metric, but it is non-vanishing
for the perturbed black string.

\bigskip

We have then found a set of physical s-wave modes on the black string in
Gauss-Bonnet theory (\ref{bs2}), which drive the instability of the
background, and therefore black strings in Gauss-Bonnet theory are unstable.

\bigskip

\section{Conclusions}

In this paper we have shown that the black strings in Gauss-Bonnet theory are
unstable under gravitational perturbations. Following the arguments in
\cite{Gregory:1994bj}, one can prove that the instability we have found cannot
be gauged away, and therefore it represents a truly physical instability.
Since the field equations are quadratic in the curvature, the linearization
around the maximally symmetric Minkowski vacua does not provide any equation
at all\footnote{As an example for how to deal with the phase space structure
of such degenerate systems see e.g. \cite{deMicheli:2012ye}.}, therefore in
order to study the perturbative properties of the solutions of Gauss-Bonnet
gravity one needs to perturb around solutions that have a non-trivial Riemann
tensor as it is the case of the black string. As mentioned above the
linearized equations around such a background are non-degenerate since all the
perturbed metric components appear in the linearized equations. In order the
black strings to be unstable, the wavelength of the perturbation along the
extended direction has to be above some minimum critical value $\lambda_{c}$.
This critical value tends to zero in the large $D$ limit in General Relativity
\cite{Emparan:2013moa}-\cite{Canfora:2009dx}, and since the large $D$ behavior
of G.R. is qualitatively similar to the one in gravity theories with a single
Lovelock term \cite{Giribet:2013wia}, one may also expect $\lambda
_{c}\rightarrow0$ as $D$ grows for the black strings and black p-branes
constructed in \cite{Giribet:2006ec}. Given the results presented in this work
it is natural to expect that the black string solution of the full
Einstein-Gauss-Bonnet theory will suffer from the Gregory-Laflamme
instability, which will induce an instability for large angular momentum in
the rotating version of the static black string constructed numerically in
\cite{Kleihaus:2009dm}.

For Einstein-Gauss-Bonnet gravity different stability analysis of black holes
have been performed in \cite{Dotti:2005sq}-\cite{Gannouji:2013eka} and it
would be interesting to extend such analysis to the whole family of black holes of
\cite{Crisostomo:2000bb}, which are the black holes in the transverse section of the black
strings we have considered here.\footnote{It would be also interesting to
extend these results to cases with more general asymptotic behaviours as the
ones e.g. \cite{Bogdanos:2009pc}-\cite{CuadrosMelgar:2010cc}.}

It's worth also to explore whether the results presented here can be extended
to all the black strings and black p-branes obtained in \cite{Giribet:2006ec},
and even to the compactifications with Einstein manifold in four dimensions
that were obtained in \cite{Canfora:2008iu} for Lovelock theories. Work along
these lines is in progress.

\bigskip

\textbf{Acknowledgments}

The authors are grateful to Marco Astorino, Fabrizio Canfora, Gustavo Dotti
and Sourya Ray for useful discussions. The authors also thank Gaston Giribet
for enlightening comments. This work has been supported by FONDECYT Regular
grants 1141073 and 1150246. A.V. appreciates the support of CONICYT fellowship 21151067. This project was also partially funded by
Proyectos CONICYT, Research Council UK (RCUK) Grant No. DPI20140053.

\end{document}